# Persistent Uncorrelated Magnetic Domains in Fe/Si Multilayers and their suppression by incorporating $^{11}B_4C$

Anton Zubayer [1,5,*], Artur Glavic[3,*], Naureen Ghafoor [1], Yuqing Ge[7], Yasmine Sassa[7], Martin Månsson[7], Andreas Suter[3], Thomas Prokscha[3], Zaher Salman[3], Wai-Tung Lee[4], Kristbjörg Anna Thórarinsdóttir [2], Arnaud Le Febvrier [1,5], Per Eklund [1,5], Jens Birch [1], Fridrik Magnus [2], Sean Langridge[6], Andrew Caruana[6], Christy Kinane[6], Fredrik Eriksson [1*]

**Affiliations**

[1]Thin Film Physics Division, Department of Physics, Chemistry, and Biology (IFM), Linköping University, SE-581 83 Linköping, Sweden

[2]Science Institute, University of Iceland; Dunhaga 3, IS-107 Reykjavik, Iceland

[3]PSI Center for Neutron and Muon Sciences, 5232 Villigen PSI, Switzerland

[4]Science Directorate, European Spallation Source, Lund 224 84, Sweden

[5]Department of Chemistry-Ångström, Uppsala University; SE-75120, Uppsala, Sweden

[6]ISIS, Harwell Science and Innovation Campus, Science and Technology Facilities Council, Rutherford Appleton Laboratory, Didcot, Oxon OX11 0QX, United Kingdom

[7]Department of Applied Physics, KTH Royal Institute of Technology, SE-106 91 Stockholm, Sweden

*Corresponding authors, e-mail: anton.zubayer@liu.se, artur.glavic@psi.ch, fredrik.eriksson@liu.se

**Abstract**

This study investigates magnetic domains in Fe/Si and Fe/Si + $B_4C$ multilayers using spin flip off-specular polarized neutron reflectometry. The results show that Fe/Si multilayers exhibit pronounced spin flip off-specular scattering originating from magnetic domains that are uncorrelated out of plane. With increasing external magnetic field the domains progressively coalesce and their magnetization rotates toward alignment with the applied field, approaching a homogeneous magnetic state at higher fields. In contrast, Fe/Si + $B_4C$ multilayers exhibit no detectable spin flip off-specular scattering already at low fields, indicating that the multilayer reaches magnetic saturation at significantly lower applied fields. The scattering patterns are interpreted using distorted wave Born approximation simulations in BornAgain, enabled by our added code for simulating magnetic domains and magnetic ordering. To further probe the magnetic behavior, low-energy $\mu^+$SR measurements were performed, representing the first $\mu^+$SR investigation of polarizing neutron optics multilayers. Together with comparison to previously reported VSM data, these measurements provide insight into the magnetic behavior across short range, medium range, and long range length scales. The results show that incorporating approximately 15 vol.% $B_4C$ makes the magnetic configuration highly responsive to external magnetic fields, with clear sensitivity to both in-plane and out-of-plane field geometries. These results show that $B_4C$ suppresses magnetic domains and spin flip off-specular scattering, improving Fe/Si coatings for neutron polarization optics in regards to off-specular scattering, and other applications requiring easy magnetic manipulation.

## Introduction

Magnetic domains in thin film multilayers are a key focus of research due to their critical role in magnetic behaviour and potential applications in spintronic devices and magnetic storage technologies. In systems like iron/silicon (Fe/Si) multilayers, the properties of magnetic domains are shaped by factors such as layer composition, magnetic moments, thickness, and interface quality. Techniques such as magneto-optical Kerr effect (MOKE) microscopy, polarized neutron reflectometry (PNR), and X-ray magnetic circular dichroism (XMCD) have been widely used to explore domain patterns and magnetic anisotropy within individual layers.[1–3] Off-specular neutron reflectivity, sensitive to lateral inhomogeneities and magnetic roughness, has emerged as a powerful tool to study domain structures and their evolution under different conditions. Recent investigations have explored how material modifications, such as introducing non-magnetic spacers or doping, influence domain formation and stability.[4–6]

The Fe/Si system and their detrimental magnetic domains are of particular importance for application in coatings for neutron polarization devices. Polarizing neutron optics play a crucial role in neutron scattering experiments by enabling precise control and analysis of neutron spin states.[7–11] High-quality polarizing optics are essential for improving the accuracy and sensitivity of measurements, which are fundamental in exploring the magnetic properties of materials in various fields such as condensed matter physics and materials science.[12] To achieve this, polarizers are constructed as multilayered systems, typically composed of alternating magnetic and non-magnetic layers, designed to selectively reflect neutrons with a desired spin orientation.[13,14] The difference in scattering contrast between spin-up and spin-down neutrons should be maximized to achieve enhanced polarization.[15,16] This difference is directly related to the magnetization of the layers. To minimize the spin-down reflectivity, the neutron contrast (SLD contrast) for the non-magnetic layer must be well matched to the spin down of the magnetic layer.

Several other factors, however, influence the polarization efficiency of such devices in negative ways. The interface morphology becomes increasingly important for multilayers used at relatively large reflection angles to retain a high reflectivity.[17,18] Furthermore, the retention of magnetic domains in the multilayer can lead to undesired spin-flip scattering that reduces polarization and therefore may require high magnetic guide fields to be suppressed.[19,20] The strength of this scattering depends on the magnetic domain sizes, the orientation of the local magnetization within a domain to the average magnetization, and the degree of correlation of domains between adjacent magnetic layers.

In previous studies, we have shown that incorporation of $^{11}B_4C$ into the Fe/Si multilayer system can improve interface quality and increase polarization of Fe/Si multilayers.[4–6,21] This is achieved by amorphization of the Fe layers thereby suppressing roughness generated by crystallite facets at the interfaces. A secondary effect is that the magnetic coercivity of the sample is practically eliminated. However, to what extent the coercivity suppression affects the spin-flip and off-specular scattering is not fully explored. This can be explored by magnetic off-specular scattering. This study aims to address this gap.

We employ off-specular neutron reflectivity with polarization analysis to investigate the magnetic domain patterns[22,23] in Fe/Si multilayers with and without $^{11}B_4C$ incorporation. The focus is on the determination of approximate domain size and vertical domain correlations, as well as to what extent the external magnetic field suppresses domain scattering. For ordinary Fe/Si multilayer we find small magnetic domains that show no correlations between adjacent magnetic layers. We explain the scattering features and simulate a simple distorted wave Born approximation (DWBA) model that can reproduce the shapes in the data. We corroborate that by reducing or eliminating crystallinity, the incorporation of $^{11}B_4C$ effectively eliminates the magnetic domains, thereby mitigating magnetic off-specular scattering which would improve the performance of polarizing neutron optics. The decrease in magnetic off-specular scattering would be advantageous for neutron optics particularly in Small Angle Neutron Scattering (SANS) beamlines, allowing for more detailed and accurate studies of material structures at the nanoscale.[24,25]

We also employ low-energy $\mu^+$SR (muon spin rotation) to probe the internal local magnetic order within Fe/Si and Fe/Si + $^{11}B_4C$ multilayers. By tuning the stopping energies of the muons, we ensured implantation predominantly within either the Fe or Si layer, allowing for detailed examination of local magnetic environments. The $\mu^+$SR measurements were conducted under varying applied magnetic fields to observe the precession behavior and modulation of local fields. This approach provided valuable complementary insights into the magnetic response and tunability of the multilayers, particularly highlighting the enhanced field responsiveness in the $^{11}B_4C$-containing samples.

## Experimental Details

Fe/Si and $^{11}B_4C$-containing Fe/Si multilayer thin films were grown using ion-assisted DC magnetron sputter deposition in an ultra-high vacuum system, maintaining a background pressure of approximately $2 \times 10^{-9}$ Torr ($2.6 \times 10^{-7}$ Pa). The details of the deposition system are described elsewhere.[26] The multilayers were deposited onto $10 \times 10 \times 0.5$ mm$^3$ Si (100) substrates with a native oxide layer, with the substrate held at ambient temperature (no intentional heating) and rotated at a constant speed of 15 rpm. The 50-mm-diameter sputtering targets were Fe (Plasmaterials, 99.95% purity), $^{11}B_4C$ (RHP Technology, 99.8% chemical purity, >90% isotopic purity), and Si (Kurt J. Lesker, 99.95% purity), operated at applied powers of 33 W, 50 W, and 20 W, respectively. Deposition rates were determined through X-ray reflectivity thickness measurements of the multilayers. The desired layer thicknesses and multilayer design were achieved using computer-controlled shutters, with opening durations tailored accordingly. During deposition, a DC substrate bias of -30 V was applied. In the case of Fe/Si + $^{11}B_4C$, each bilayer was formed by co-sputtering $^{11}B_4C$ with Fe, followed by the co-deposition of $^{11}B_4C$ with Si. The deposition rates for both Fe + $^{11}B_4C$ and Si + $^{11}B_4C$ were approximately 0.5 Å/s. The amount of $^{11}B_4C$ in the Fe and Si layers are 15 vol.% (or 15 atomic % of $^{11}B$ + C). The 2 samples made for the XRR, XRD and PNR were Fe/Si and Fe/Si + $^{11}B_4C$ with 100 Å period thickness, a thickness ratio of 0.5 and 10 periods. While a second batch of the same materials Fe/Si and Fe/Si + $^{11}B_4C$ but with a period thickness of 500 Å and 2 periods, were made for the $\mu^+SR$ experiments.

X-ray reflectivity (XRR) analysis was performed using a Panalytical Empyrean diffractometer configured in parallel beam geometry. The instrument utilized a line-focused Cu-Kα (wavelength 1.54 Å) anode source which was operated at 45 kV and 40 mA. To condition the incident beam and limit the X-ray spot size on the sample, a parabolic X-ray mirror and a ½° divergence slit were employed in the beam path. On the diffracted beam side, a parallel plate collimator with a 0.27° collimator slit was used, followed by a PIXcel detector operating in open detector mode. To analyze the XRR data, the multilayer period thickness and interface roughness were determined through a fitting process using the GenX software (v3.6.24).[27]

X-ray diffraction (XRD) was conducted using a Panalytical X'Pert diffractometer configured in Bragg-Brentano geometry, with a Cu-Kα (wavelength 1.54 Å). The setup included a Bragg-Brentano HD incident beam optics module equipped with a ½° divergence slit and a ½° anti-scatter slit. On the secondary optics side, a 5 mm anti-scatter slit was paired with an X'celerator detector operating in scanning line mode. Diffraction measurements were taken over a 2θ scanning range of 20° to 90°.

The magnetic properties of the samples were examined using vibrating sample magnetometry (VSM) on a quantum design PPMS,[28] in a longitudinal geometry at room temperature. Magnetic hysteresis curves were recorded over a field range of -20 mT to 20 mT, allowing for the visualization of the coercive field of the samples.

Off-specular neutron reflectivity with polarization analysis was performed at the POLREF beamline at the ISIS neutron and Muon Source in the UK.[29] The PORLEF beamline is equipped with a front (incident beam) neutron polarizer and back end (reflected beam) neutron analyzer setups allowing the incident neutron spin eigenstate to be polarized and the reflected neutron spin eigenstate to be analyzed. This allows both the Non-spin Flip (NSF) and Spin Flip (SF) to be measured separately.[30–32] The combination of NSF and SF measurements allows the magnitude and orientation of magnetization relative to the applied magnetic field to be measured as a function of depth through the sample. It is in essence a depth-dependent in-plane vector magnetometer. The off-specular analysis allows in-plane information to be measured within the length scale of a few 100 nm to 30 µm [33–36] giving information on domain sizes, magnetic magnitude and orientation. Smaller domain sizes are limited by the accessible $q_x$-range while the neutron coherence (given by resolution) limits the largest accessible sizes. The samples were mounted in a room temperature bore GMW magnet capable of applying a uniform field of ±0.7T, which also defines the quantization axis for the neutron polarization. The efficiencies of the polarization devices were corrected for with reference measurements following the method described by Wildes et al.[37]

The BornAgain software[38] was employed to simulate off-specular neutron reflectivity in all four spin states using DWBA, thus including refraction and reflection effects. The simulations were iteratively modified to find a suitable match with the data itself, primarily in regards to shape. Specular profiles were extracted from the off-specular simulations. The simulations allowed for the extraction of key structural and magnetic parameters of the Fe/Si and Fe/Si + $^{11}B_4C$ multilayers. Specifically, we obtained the layer thicknesses, interface roughnesses, magnetic domain sizes, and the orientation angles of the magnetic moments within the domains. The ability to model both the specular and off-specular scattering provided a comprehensive understanding of the magnetic structure and interlayer properties. The novel code enabling simulation of magnetic off specular scattering in BornAgain, allowing magnetic structures to be modeled within the framework for the first time, is provided in the supplementary material in detail with explanations of workflow and code.

The Muon spin rotation ($\mu SR$) experiment is performed at the low-energy $\mu SR$ spectrometer (LE- $\mu SR$)[39] at the Swiss Muon Source (SµS), Paul Scherrer Institute, Villigen, Switzerland. The data is analysed with the software musrfit.[40] The two samples were measured with external applied fields of 0, 1, 5 and 30 mT. The LE-$\mu SR$ time spectra are fitted with:

$$A_0 P_{TF}(t) = A_{B_0} \exp(-\lambda_{B_0} t) \cos(B_0 \gamma_\mu t) + A_S \exp(-\lambda_S t) + A_F \exp(-\lambda_F t) \quad (1)$$

where the $A_0, A_{B_0}, A_S,$ and $A_F$ are the total initial asymmetry, the initial asymmetry related to the local fields, and the slow and fast exponential relaxations, respectively. The $\lambda_i$s are the corresponding relaxation rates, the $B_0$ represents the average effective local fields muons experience in the sample as a result of the external fields combined with the internal fields from the magnetic order of the sample, and the $\gamma_\mu = 2\pi \times 0.01355\ MHz/G$ is the gyromagnetic ratio of the muon. The $B_{ext}$ = 0 mT measurements were fitted assuming $A_{B0}$ = 0. The 2 samples used for $\mu SR$ was specifically made thicker (500 Å period thickness) in order to control if the implantation depth is primarily in the Fe or Si layers.

## Results and Discussion

Two multilayers, Fe/Si and Fe/Si + $^{11}B_4C$ (co-sputter $^{11}B_4C$ with Fe and Si), were investigated. Each was designed to have a periodic thickness of Λ = 100 Å, a layer thickness ratio of Γ = 0.5, and a total of N = 10 periods. These multilayer parameters were selected to produce sufficient signal for the experiment while avoiding effects due to accumulated roughness which would complicate the modelling.

Previously published XRR measurements[4] and GenX3 fitting[27] determined the multilayer parameters. As reported earlier[4] and shown again in Figure 1(a), the Fe/Si + $^{11}B_4C$ multilayer exhibits higher reflectivity up to the 7th order Bragg peak compared to Fe/Si. The fitted interface width is smaller for Fe/Si + $^{11}B_4C$ (σ = 7.8 Å) than for Fe/Si (σ = 11.5 Å). XRD results in Figure 1(b) confirm that the $^{11}B_4C$-containing multilayer is X-ray amorphous, whereas Fe/Si shows crystalline Fe and possible iron silicide phases. As discussed previously[4], the enhanced reflectivity despite reduced SLD contrast is attributed to $^{11}B_4C$-induced amorphization, leading to smoother interfaces and reduced intermixing. The reduced interface width explains the higher X-ray and neutron reflectivities[4].

Consistent with earlier results for thinner layers[4], the present multilayer with $\Lambda$ = 100 Å shows negligible remanent magnetization and no coercivity in VSM for the $^{11}B_4C$-containing sample, in contrast to Fe/Si (Figure 1(c)). The absence of coercivity is linked to the amorphous structure and the resulting modification of magnetic domain formation due to the lack of crystalline anisotropy[19,41]. Figure 1(c) also marks the field values used for the neutron off-specular measurements in this study: 2 mT and 12 mT for both samples, and 700 mT for Fe/Si.

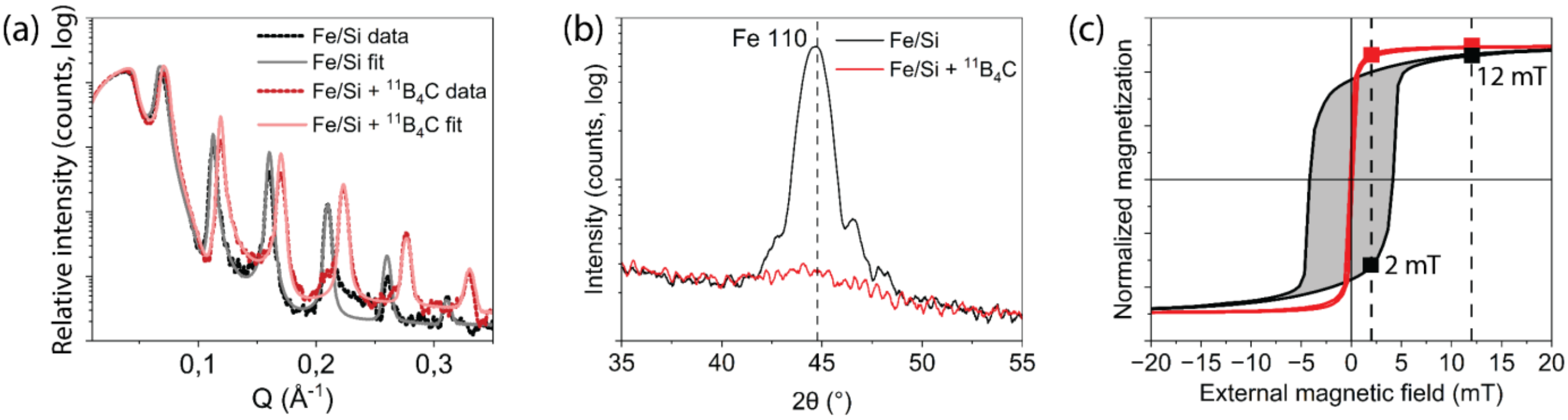

*Figure 1. (a) X-ray reflectivity (XRR) with corresponding fits for the Fe/Si (black) and Fe/Si + $^{11}B_4C$ (red) samples. (b) X-ray diffraction (XRD) focusing on the angular region of 35-55°. (c) Vibrating sample magnetometry (VSM) with external fields between -20 and 20 mT. The magnetization is normalized to the saturation value. The red and black squared points represent the external fields used for measuring PNR for Fe/Si and Fe/Si + $^{11}B_4C$. The data in a) and c) are from our previously published work, Ref*[4].

Magnetic off-specular scattering arises due to the presence of magnetic domains within the Fe layers. To distinguish structural off-specular scattering from magnetic off-specular scattering, one must observe the behavior of the spin of scattered neutrons. If the off-specular scattering is magnetic in nature, the spin of the neutrons will flip after interacting with a magnetic domain (perpendicular deviation from the average moment) or a difference for spin-up and spin-down neutrons is observed (longitudinal deviation from average moment). This occurs because the magnetic moment within the domain is oriented differently from the applied external magnetic field and the layer averaged magnetization. Neutrons enter the multilayer structure and, upon encountering a magnetic domain within the Fe layer, may change their trajectory (scatter). The interaction with the magnetic domain also causes the neutrons to flip their spin.

In a multilayer structure the scattering is a complex combination of interference effects from waves reflected on all interfaces within the sample and the scattering within each layer that effect any of these waves. All these effects are considered in DWBA models. A qualitative visual explanation of the observed scattering patterns is depicted in Figure 2.

If the magnetic domains of adjacent magnetic layers are correlated (i.e. a magnetic coupling between layers exists) then the scattering amplitudes have a fixed phase relation that is equivalent with the relation between reflections from interfaces with the same distance. This leads to a strong off-specular signal at the same $q_z$-position as the multilayer Bragg-peak (a so-called Bragg-sheet depicted in Figure 2d). The fact that the optical parameters for spin-up and spin-down are different leads to an asymmetry between off-specular scattering to the different sides of the specular line for spin-flip channels. This asymmetry is inverted between up-down and down-up channels, which can be understood as an inversion of beam direction.

In the opposite case with no correlations between the magnetic domains of different layers, the scattering is added incoherently and no enhancement at the $q_z$ of the Bragg-peak is observed. The incoming and outgoing beams, however, are still reflected on the sample interfaces. This means that there is not only a direct scattering from neutrons entering the sample towards the detector, the diffuse intensity which is strongest at low q close to the critical edge of total reflection (blue in Figure 2c and d), but also a possibility to reflect before or after the scattering process. Ideally for a neutron polarizing Fe/Si system the spin-down contrast is negligeable, and ideally no Bragg-reflection is observed for spin-down neutrons. If there are magnetic domains, the spin-up incident beam can make a Bragg-reflection and then scatter with spin-flip before leaving the sample freely (Figure 2a). Similarly, spin-down incident beam encounters the opposite effect, entering the sample freely before scattering and then being Bragg-reflected (Figure 2b). These situations lead to peculiar scattering patterns with triangular shapes that are restricted by the critical angle on one side and a diagonal with $p_i/p_f = p_{Bragg}$ (green/red area marked with arrows in Figure 2c). However, Figure 2c is the case of uncorrelated magnetic domains while 2d for correlated magnetic domains.

In this study, we employ a coordinate system wherein the out-of-plane momentum vector components $p_i$ (initial) and $p_f$ (final) are used as they match well to the observed off-specular scattering.[35,42] The out-of-plane momentum transfer, denoted as $q_z$, is the sum of $p_i$ and $p_f$, capturing correlations normal to the sample's surface. We define $p_{Bragg} = p_i = p_f = q_z/2$ at the superlattice specular Bragg peak. This choice facilitates the characterization of specular and off-specular scattering components, ensuring accurate quantification of scattering phenomena in relation to the Bragg condition.

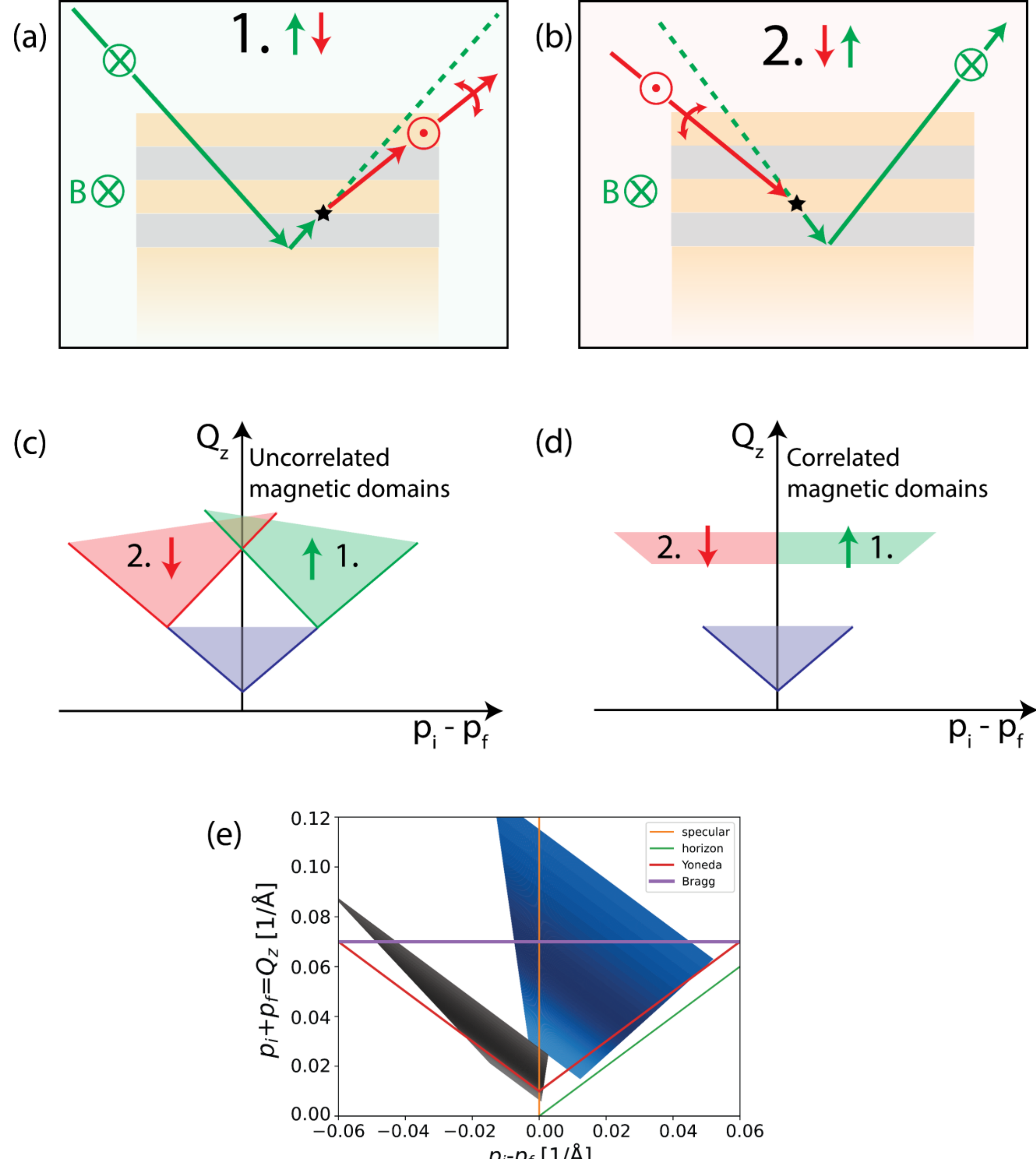

*Figure 2. Sketch of the reason and pathway of spin-flip off-specular scattering caused by magnetic domains; spin-up (a) or spin-down (b) neutrons entering a multilayer with magnetic domains. The applied external field, B, points into the paper. The star represents a point where the neutron is interacting with a magnetic domain in an Fe layer. In both cases the spin-up beam has enhanced reflectivity at the Bragg-peak. The bottom figures show sketches of the resulting spin-flip scattering patterns for vertically uncorrelated (c) and correlated (d) magnetic domains. In magnetic off-specular neutron scattering, $p_i$ and $p_f$ represent the components of the neutron wavevector along the surface normal for the incident and scattered beams, respectively, and are crucial for determining the perpendicular momentum transfer $Q_z$, which is sensitive to variations in magnetic structures like domains or spin textures. In the $Q_z$ vs $p_i$–$p_f$ representation the specular reflectivity is vertical and Yoneda lines diagonal. (e) shows the two cuts (blue and black) that the neutron measurements gathered data. The orange, purple, red and green lines indicates the specular, horizon, Yoneda and Bragg lines.*

Figure 3 shows the data and corresponding simulations from the spin-flip off-specular neutron reflectivity measurements for Fe/Si and Fe/Si + $^{11}B_4C$ to extract information on magnetic domain sizes and the degree of their rotation relative to the external magnetic field. The non-spin-flip results for the two samples and additionally 700 mT applied field for Fe/Si sample are shown in the supplemental information. The lower left areas for all the Fe/Si + $^{11}B_4C$ measurements were not measured due to time limitation, but also due to no need to measure because of the very low signal of off-specular intensity in general. Measurements were performed at external fields of 2 mT (Figure 3 (a-b) and 12 mT (Figure 3 (c-d). The corresponding can be seen for Fe/Si + $^{11}B_4C$ in Figure 3(i-j) and (k-l). Additional measurements made at 700 mT for the Fe/Si sample are shown in S3(s-t). Whereas the experimental data already shows visual differences between the multilayers, the corresponding simulations, to mimic the shapes seen in the data, provide some quantitative information on domain sizes and rotation angles. The key off-specular intensities of interest here are the spin-flip intensities (i.e. all the plots seen in Figure 3), as they reveal the presence and behaviour of magnetic domains. When comparing the two multilayers, it is evident that the Fe/Si + $^{11}B_4C$ exhibits less, if any, off-specular scattering across any applied external magnetic fields, as seen in Figure 3(i-p). In contrast, the pure Fe/Si shows pronounced off-specular intensities, especially at 2 mT. This difference is consistent with the VSM results in Figure 1(c), which indicate that the Fe/Si is not magnetically saturated at 2 mT, while the $^{11}B_4C$ containing multilayer is nearly fully saturated. Notably, at 2 mT, the Fe/Si remains in a state of negative magnetization, corroborating the VSM results. Further, at 12 mT, where the external field has passed the coercive field, the Fe/Si still shows off-specular scattering. This observation is consistent with the observed incomplete saturation of magnetization in Figure 1(c). Figure 1(e) shows cuts of the sketch indicating where the neutron data were collected. These cuts arise from the incident angles and the instrument geometry. The blue region indicates that the specular, Bragg, and Yoneda intensities are visible, making this range sufficient for the present study. Especially due to its symmetrical nature.

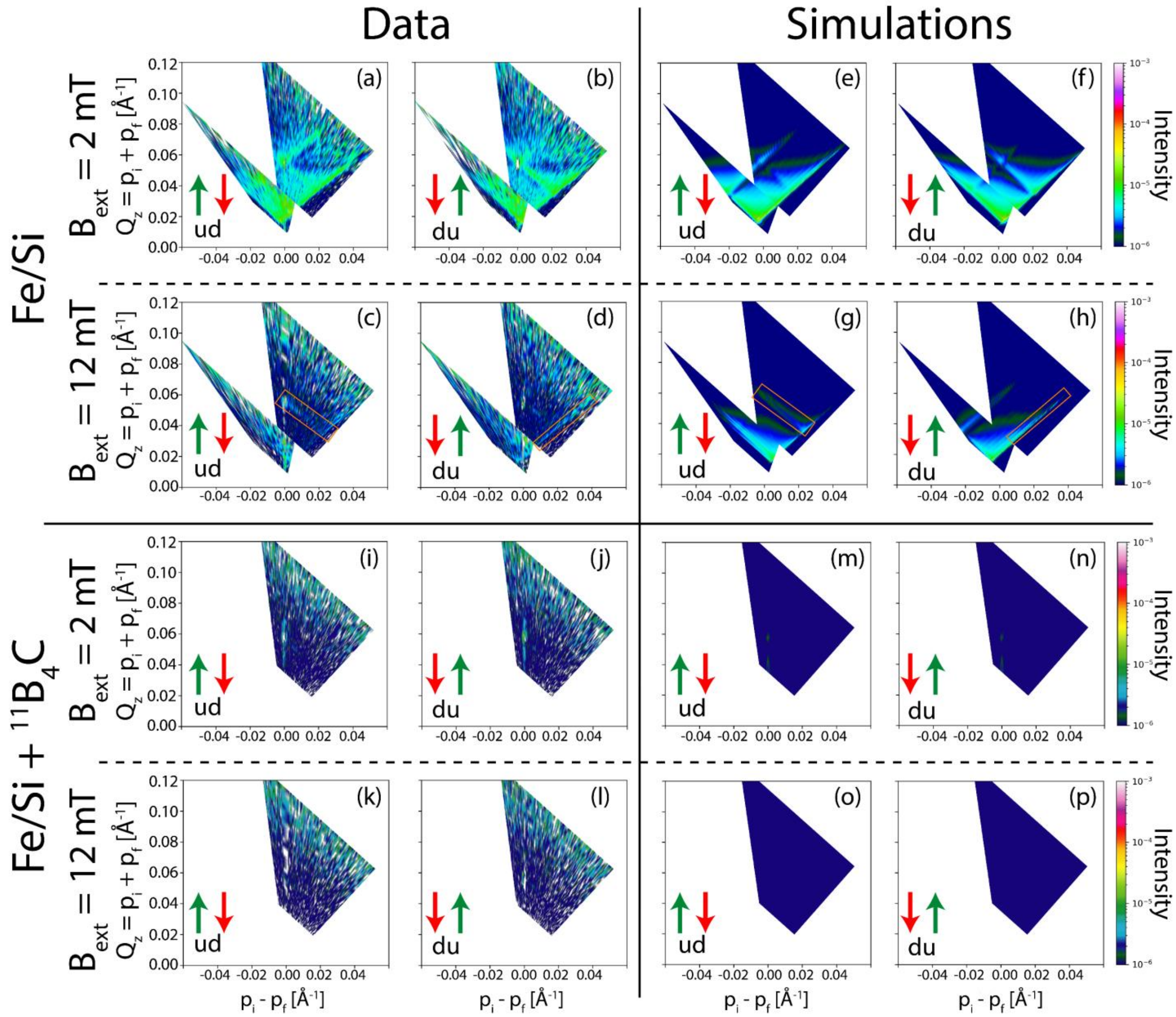

*Figure 3. Spin-flip off-specular polarized neutron reflectivity (PNR) of Fe/Si. (a-b) showing the raw data for the 2 mT and (c-d) for 12 mT. While (e-f) shows the corresponding spin-flip off-specular simulations, to mimic the features seen in the data, using BornAgain for 2 mT and (g-h) for 12 mT. (i-j) and (k-l) are the spin-flip off-specular PNR of Fe/Si + $^{11}B_4C$ measured at 2 and 12 mT external field respectively, while (m-n) and (o-p) shows the corresponding spin-flip off-specular simulations for 2 mT and (g-h) for 12 mT. All panels use a colour plot which indicates the neutron intensities in a log scale. Orange boxes surrounding certain parts of the plots in (c-d) and (g-h) highlights the similarities between the data and simulations.*

At 2 mT, the Fe/Si + $^{11}B_4C$ multilayer (Figure S4(a-d)) exhibits significantly less magnetic, and even non-magnetic, scattering compared to the pure Fe/Si multilayer at 12 mT (Figure S4(i-l)) and arguably even 700 mT (Figure S4(q-t)). This indicates that at very low field of 2 mT the $^{11}B_4C$ containing multilayer had substantially less, if any, magnetic domain-related scattering compared to the pure Fe/Si multilayer and would therefore minimize the off-specular scattering. The elimination of crystallinity in Fe/Si + $^{11}B_4C$ leading to the elimination of coercivity and more easy magnetic saturation, is here confirmed directly causing the absence of magnetic domains. It is this lack of domains effectively reducing the off-specular scattering, as observed in the neutron reflectivity data. The amorphization caused by B, in such as FeB and FeCoB, has been previously reported and has been heavily studied in terms of magnetization and structure.[43–46] However, investigations if magnetic domains in Fe/Si multilayers, which is the most common polarizing neutron optics material combination, are correlated or not and investigating if the magnetic domains are supressed or not when $^{11}B_4C$ is added is unique.

The simulations in Figure 3(e-f), for 2mT, use a model of uncorrelated magnetic domains in the Fe/Si sample, with domain sizes of 250 nm and an angle of 130° relative to the external field. In Figure 3(g-h), at 12 mT, the domain size increases to 420 nm, and the angle decreases to 20°, indicating a shift towards better alignment with the external magnetic field as the external field strength increases. A schematic illustration based on these conclusions is visualized on the right side in Figure 4. The increase of magnetic domain sizes suggest that the magnetic domains have coalesced as the external field was increased from 2 mT to 12 mT. The off-specular spin-flip scattering simulation at 700 mT, in Figure S3(w-x) the off-specular scattering in the Fe/Si sample is absent, indicating that the magnetic domains have fully aligned with the external magnetic field, effectively eliminating the scattering caused by domain misalignment. It should be noted that a magnetic field of 700 mT is significantly higher than the 20 mT typically used to operate polarizing neutron optics, suggesting it impractical for most applications despite its ability to eliminate off-specular scattering caused by magnetic domains by simply having a large enough external field. Especially considering that a polarizing neutron optics mirror typically contains more than 1000 periods, rather than the 10 used in this study, which would correspond to spin flip off specular scattering originating from two orders of magnitude more interfaces. Therefore, if a much lower field of only 2 mT can eliminate magnetic domains for Fe/Si + $^{11}B_4C$, this is a far more suitable option for polarizing neutron optics from the perspective of decreasing off-specular scattering.

One-dimensional scans along the $Q_z$direction were extracted from the off-specular measurements to obtain the specular reflectivities (see Supplementary). For the Fe/Si sample, the spin-down reflectivity exceeds the spin-up reflectivity at 2 mT, indicating that the average magnetic moment remains partially aligned opposite to the external field. At higher fields the polarization improves as the magnetization approaches saturation, whereas the Fe/Si + $^{11}B_4C$ sample shows nearly identical reflectivities already at low fields, consistent with its faster magnetic saturation

A different model than the uncorrelated cylindric magnetic domains was also evaluated, to simulate the 4-spin state off-specular PNR. As shown in the Supplementary Information, the various correlated magnetic domain models fail to match the experimental data. Adjusting the domain size or shape and orientation did further diverge from experimental observations, too. We therefore conclude that the Fe/Si multilayer exhibit uncorrelated magnetic domains with a typical size around 200 nm under a external field of 2 mT and as the external field increases, the domains coalesce and start to align with the external field direction.

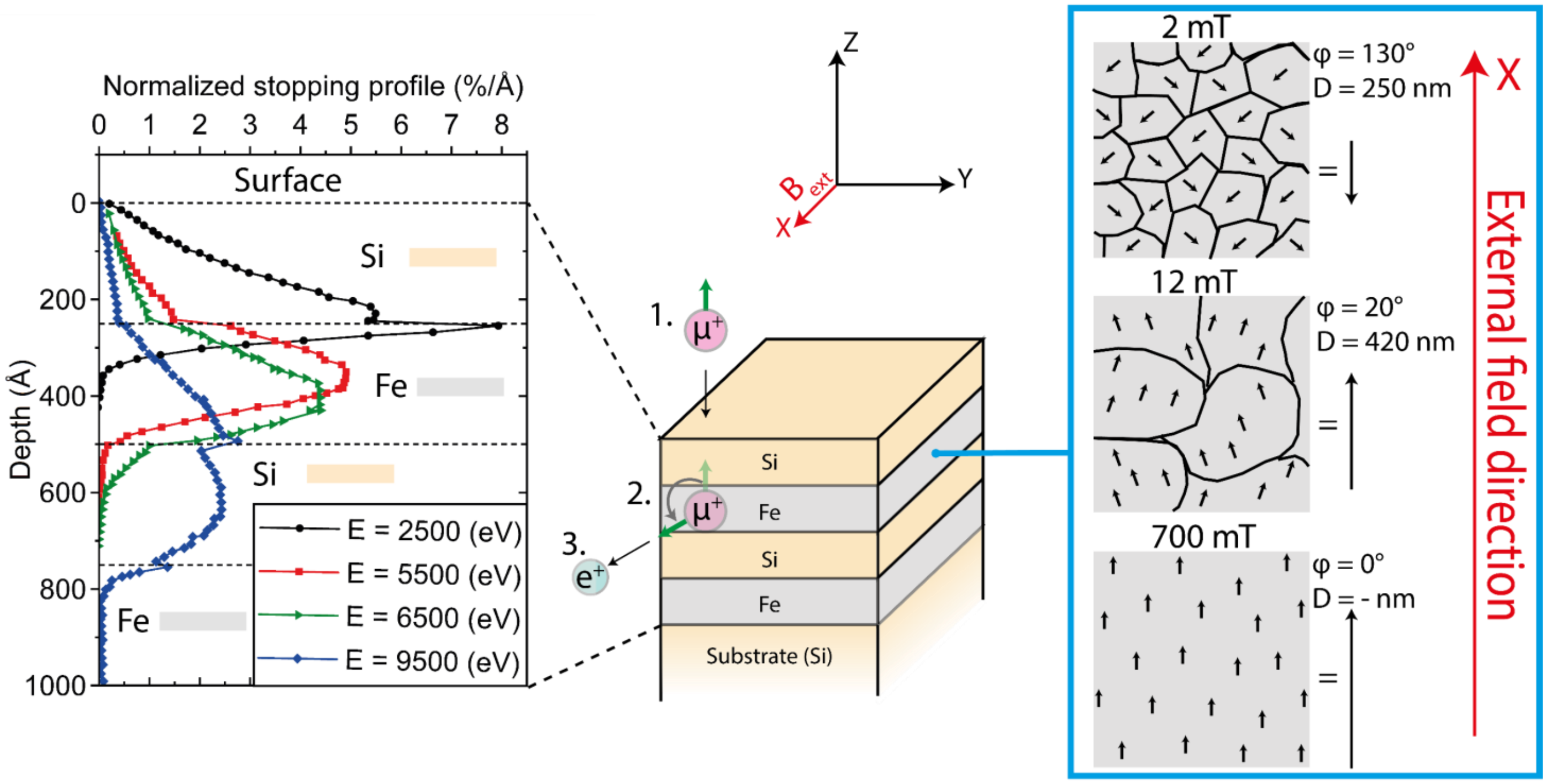

*Figure 4. Schematic of the multilayer (center) along with the muon* $(\mu^+)$ *traversing towards it (1), implanting and precessing (2) and decaying into a positron* $(e^+)$, *neutrino* $(\nu_e)$ *and antineutrino* $(\bar{\nu}_e)$ *where the positron will eject in the direction of the muons last spin direction before decaying (3), before finally being captured by the detectors. The muon stopping profile is seen on the left where the muons are mainly implanted in different layers depending on their initial energy. The magnetic domains seen on the right (black domain boundaries) with their corresponding local alignment in relation to the external field (arrows inside the domains). The arrows to the right of the schematic domains show the net magnetization and the alignment with respect to the external field. The domain sketches, angle of canting* $(\varphi)$ *and domain sizes* $(D)$ *are results from the magnetic off-specular measurements and simulations.*

To investigate the internal, local, magnetic order in Fe/Si and Fe/Si + $^{11}B_4C$ multilayers, low energy $\mu^+SR$ was performed on multilayers with the period thicknesses of 500 Å instead of 100 Å. These thicknesses were chosen due to the muon implanting stopping energies needed to make sure most of the muons would implant predominantly only in the first Fe or Si layer from the top. The energies of the muons are tuned to be mostly implanted in Fe layers and Si layers with energies of 6 keV and 9.5 keV, respectively. The muon stopping energy needed to implant in Fe/Si and Fe/Si + $^{11}B_4C$ multilayers are negligibly different. The external magnetic fields are applied perpendicular (transverse) to the initial spins of the muons.

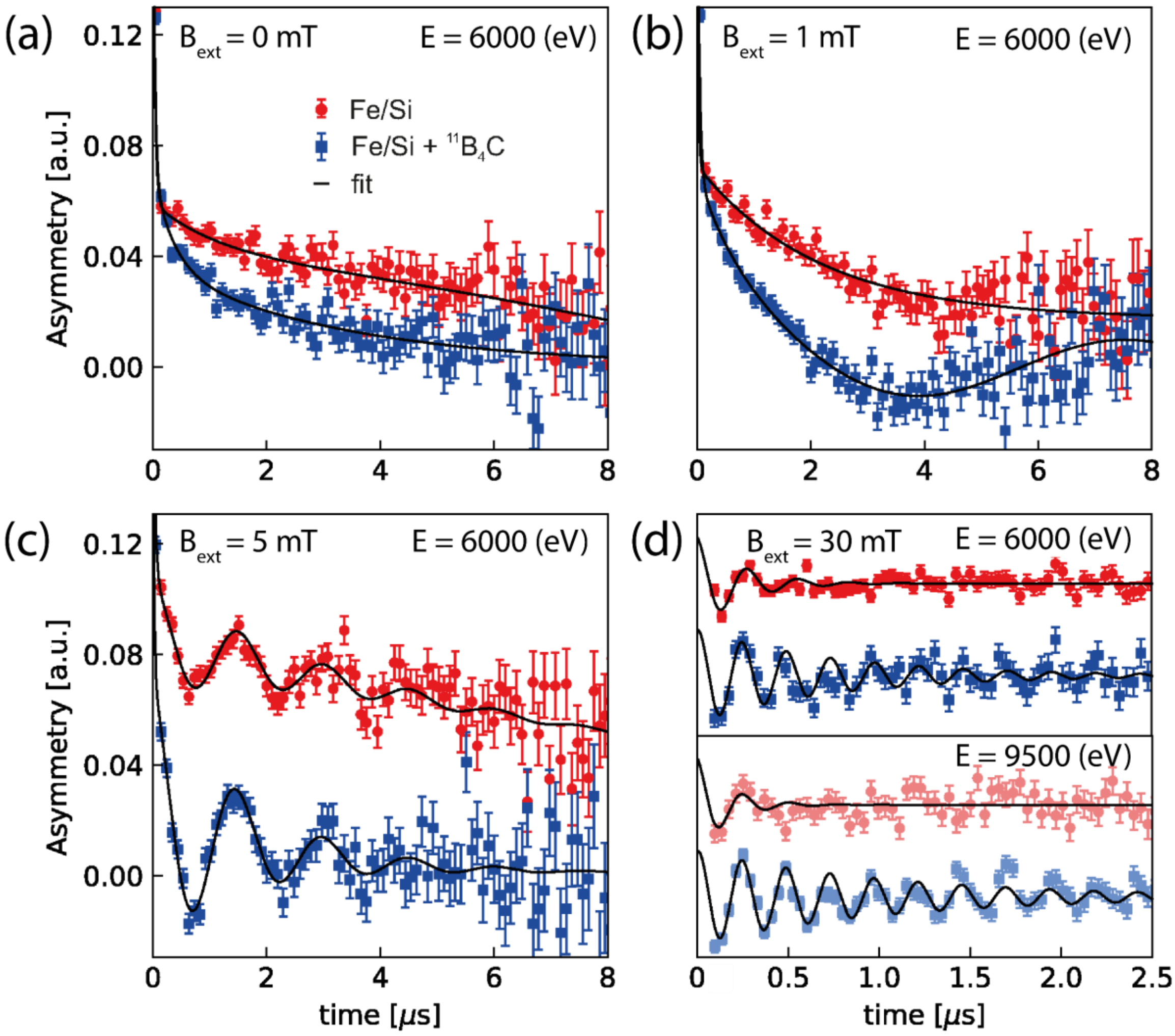

*Figure 5. LE- μSR time spectra obtained at various external fields applied parallel to the sample surface, at (a) 0 mT (b) 1 mT, (c) 5 mT, and (d) 30 mT. Spectra shown in (d) are the oscillation components normalized by the initial asymmetries of oscillation and shifted in the y-axis for clarity of view. The energy of 6000 eV corresponds to the majority of the muons implanted into the Fe layer, while with the 9500 keV muons are mostly in the Si layer. The sample has been saturated in an applied field of 55 mT in the opposite direction before (a-c) and (d) were carried out, separately. All measurements were carried out at room temperature. The red and blue data points represents the Fe/Si and Fe/Si + $^{11}B_4C$ sample respectively. The solid lines are fits corresponding to Equation (1).*

As seen in Figure 5(a-b), an applied field of 1 mT is large enough to induce a local field visibly manifesting the muon spin precession in the Fe/Si + $^{11}B_4C$ sample thereby a cosine oscillation can be seen in the spectrum, while no substantial oscillation of comparable magnitude can be seen in the Fe/Si sample. The presence of the oscillation with a frequency corresponding to the external field in the Fe/Si + $^{11}B_4C$ sample indicates that the local magnetic environment inside the Fe layer can be well modulated by an external field of 1 mT. Upon increasing the applied field, such kind of oscillations start to be present in both samples. However, the oscillations are much less relaxed and long-persisting in the $^{11}B_4C$ containing layers, especially at an external field of 30 mT with relaxation rates $\lambda_{B_0} = 2.1(8)/\mu s$ for the Fe/Si + $^{11}B_4C$, but a $\lambda_{B_0} = 5.1(9)/\mu s$ for the Fe/Si. In the spectra, the oscillations of the best fits persist for more than 2.5 $\mu s$ in the Fe/Si + $^{11}B_4C$ sample, while only up to 2 cycles of the cosine functions are present in the Fe/Si. $\mu^+$SR serves as a local probe, providing unique insights into the uniformity of the local magnetic environment at the muon stopping site. While bulk-averaged (VSM) and long-range (PNR) techniques highlight the macroscopic effects of $^{11}B_4C$ inclusion, the LEM results reveal that this effect persists even at the short range muon length-scale. Here we refer the muon length-scale using LE-$\mu^+$SR to a few nanometres.[47] The observed oscillating asymmetries in the Fe/Si + $^{11}B_4C$ sample, along with the possible absence of multiple frequencies in the spectra, suggest no phase separation at the local level and confirm the single-domain-like behavior. The reduced relaxation rates and longer-persisting oscillations in the $^{11}B_4C$-containing sample further indicate that the inclusion of $^{11}B_4C$ enhances the uniformity of local magnetic fields. This consistency across bulk, long-range, and local probes demonstrates that the favorable magnetic properties of $^{11}B_4C$-containing multilayers arise from its stabilizing effect on the local magnetic environment.

## Conclusions

We investigated the magnetic domain structure in Fe/Si and Fe/Si + $^{11}B_4C$ multilayers using spin flip off specular polarized neutron reflectometry, supported by low energy $\mu^+$SR and comparison with previously reported VSM data. This is the first $\mu^+$SR study of polarizing neutron optics multilayers and the first work to combine probes of short range, medium range, and long range magnetic order in such systems. In Fe Si, the spin flip off specular signal originates from magnetic domains that are uncorrelated out of plane. At low external fields of 2 and 12 mT, pronounced spin flip off specular scattering indicates the presence of magnetic domains whose magnetization is canted relative to the applied field. With increasing external field the domains progressively coalesce and their magnetization rotates toward alignment with the field, while at 700 mT the off specular scattering disappears, indicating a homogeneous magnetic state. In Fe Si + $^{11}B_4C$, no spin flip off specular scattering is observed already at the low field of 2 mT, indicating that the multilayer reaches magnetic saturation at very low applied fields. The magnetic off specular scattering was interpreted using distorted wave Born approximation simulations in BornAgain, enabled by our added code that allows BornAgain to simulate magnetic domains and magnetic ordering. The simulated scattering patterns reproduce the main features and shape of the experimental data and can also be used by others to model magnetic thin film systems.

Combining PNR, $\mu^+$SR, and the VSM comparison shows that incorporating about 15 vol.% $^{11}B_4C$ makes the magnetic configuration highly responsive to external magnetic fields, with clear sensitivity for both in plane and out of plane field geometries. Such behavior is particularly beneficial for polarizing neutron optics, where magnetic multilayers must be reliably saturated at low fields, and may also be advantageous for other magnetic thin film applications where facile magnetic manipulation is desired.

## Acknowledgements

We would like to thank the ISIS neutron source of the provision of beamtime (RB2310605). The ISIS data is available at the following DOI https://doi.org/10.5286/ISIS.E.RB2310605-1. This work was supported by the Swedish Research Council (VR), grant number 2019-04837_VR. A.Z acknowledge the grant 2022-D-03 from the Hans Werthén Foundation, the grant from Royal Swedish Academy of Sciences Physics grant, PH2022-0029, the Lars Hiertas Minne foundation grant FO2022-0273 and the Längmanska Kulturfonden grant BA23-1664. MM acknowledge funding from the Swedish Research Council, VR (Dnr. 2021-06157 and Dnr. 2022-03936), the Swedish Foundation for Strategic Research (SSF) within the Swedish national graduate school in neutron scattering (SwedNess), and the Carl Tryggers Foundation for Scientific Research (CTS-22:2374). Y.S. acknowledge funding from the Knut and Alice Wallenberg Foundation through the grant 2021.0150. Y.G. acknowledge funding from the KTH Royal Institute of Technology, Materials Platform. The μSR experiments were performed at the Swiss Muon Source SμS, Paul Scherrer Institute, Villigen, Switzerland. We highly appreciate the great technical and scientific support from the local staff.

## References

1. Yu, C., Jiang, H., Shen, L., Flanders, P. & Mankey, G. The magnetic anisotropy and domain structure of permalloy antidot arrays. *J. Appl. Phys.* **87**, 6322–6328 (2000).

2. Kudo, K. *et al.* Simulations of magnetic domain patterns on the surface of Co/Ni multilayers. *Surf. Interface Anal.* **46**, 962–969 (2014).

3. Welp, U., Vlasko-Vlasov, V., Liu, X., Furdyna, J. & Wojtowicz, T. Magnetic domain structure and magnetic anisotropy in Ga1-xMn(x)As. *Phys. Rev. Lett.* **90**, 167206 (2003).

4. Zubayer, A. *et al.* Reflective, polarizing, and magnetically soft amorphous neutron optics with 11 B-enriched B 4 C. *Sci. Adv.* **10**, 1–7 (2024).

5. Zubayer, A. *et al.* Magnetic hysteresis control in thin film Fe/Si multilayers by incorporation of B4C. *Mater. Des.* **257**, 114387 (2025).

6. Zubayer, A. *et al.* Diminished spin-flip reflectivity in stacked multilayers with varying period thicknesses of Fe / Si by incorporating 11B4C. *Mater. Today Adv.* **26**, 100578 (2025).

7. Kulda, J. *et al.* Neutron optics of the ILL high-flux polarized neutron three-axis spectrometer IN20B. *Proc. SPIE* (2001) doi:10.1117/12.448071.

8. Gubarev, M., Ramsey, B. & Mildner, D. Grazing-Incidence Neutron Optics based on Wolter Geometries. *Proc. SPIE* (2008).

9. Shimizu, H. Applications of Neutron Optics. *HAMON* (1991) doi:10.5611/HAMON.14.63.

10. Krist, T. Trevor Hicks and solid state neutron optics. *J. Phys. Condens. Matter* (2009) doi:10.1088/0953-8984/21/12/124208.

11. Mildner, D., Chen, H., Downing, R. G., Benenson, R. E. & Glinka, C. Low-resolution small-angle scattering using neutron focusing optics. *J. Phys. IV* (1993) doi:10.1051/JP4:1993889.

12. Garlea, O. *et al.* VERDI: VERsatile DIffractometer with wide-angle polarization analysis for magnetic structure studies in powders and single crystals. *Rev. Sci. Instrum.* (2022) doi:10.1063/5.0090919.

13. Peskov, B., Pleshanov, N., Pusenkov, V. & Syromyatnikov, V. NEUTRON OPTICAL DEVICES OF PNPI. *Proc. PNPI.*

14. Schreyer, A. *et al.* Spin polarized neutron reflectivity study of a Co/Cu superlattice. *J. Appl. Phys.* (1993) doi:10.1063/1.353958.

15. Majkrzak, C. F. & Berk, N. F. Neutron flux enhancement using spin-dependent interaction potentials. *Phys. Rev. B* **40**, 371 (1989).

16. Heinemann, A. & Wiedenmann, A. Benefits of polarized small-angle neutron scattering on magnetic nanometer scale structure modeling. *J. Appl. Crystallogr.* **36**, 1296–1302 (2003).

17. Eymery, J., Hartmann, J. & Baumbach, G. Interface dilution and morphology of CdTe/MnTe superlattices studied by small- and large-angle x-ray scattering. *J. Appl. Phys.* **87**, 3723–3731 (2000).

18. Fragneto, G. & Menelle, A. Progress in neutron reflectometry instrumentation. *Eur. Phys. J. Plus* **126**, 1–3 (2011).

19. Paul, A., Kentzinger, E., Rücker, U., Bürgler, D. & Brückel, T. Field-dependent magnetic domain structure in antiferromagnetically coupled multilayers by polarized neutron scattering. *Phys. Rev. B* **73**, 94441 (2006).

20. Sato, T. *et al.* Neutron-depolarization analysis and small-angle neutron-scattering studies of the reentrant spin glass Ni77Mn23. *Phys. Rev. B* **48**, 6074 (1993).

21. Zubayer, A. *et al.* The Role of 11B4C Interlayers in Enhancing Fe / Si Multilayer Performance for Polarized Neutron Mirrors. (2025) doi:10.1021/acs.jpcc.5c00068.

22. Maruyama, R. *et al.* Study of the in-plane magnetic structure of a layered system using polarized neutron scattering under grazing incidence geometry. *Nucl. Instruments Methods Phys. Res. Sect. A Accel. Spectrometers, Detect. Assoc. Equip.* **819**, 37–53 (2016).

23. Maruyama, R. *et al.* Accessible length scale of the in-plane structure in polarized neutron off-specular and grazing-incidence small-angle scattering measurements. *J. Phys. Conf. Ser.* **862**, (2017).

24. Alefeld, B., Fabian, H. & Springer, T. Recent Studies In Neutron Optics, In Particular For Small-Angle Neutron Scattering. in *Proceedings of SPIE* vol. 1149 9–14 (1989).

25. Mildner, D., Hammouda, B. & Kline, S. A refractive focusing lens system for small-angle neutron scattering. *J. Appl. Crystallogr.* **38**, 933–939 (2005).

26. le Febvrier, A. *et al.* An upgraded ultra-high vacuum magnetron-sputtering system for high-versatility and software-controlled deposition. *Vacuum* **187**, (2021).

27. Glavic, A. & Björck, M. GenX 3: The latest generation of an established tool. *J. Appl. Crystallogr.* **55**, 1063–1071 (2022).

28. Design, Q. Physical Property Measurement System (PPMS): EverCool II cryogen-free refrigerator. (2024).

29. Polref Instrument Overview.

30. Moon, R. M., Riste, T. & Koehler, W. C. Polarization analysis of thermal-neutron scattering. *Phys. Rev.* **181**, 920–931 (1969).

31. Blundell, S. J. & Schofield, A. J. Neutron polarization analysis in magnetism. *J. Magn. Magn. Mater.* **121**, 185–188 (1993).

32. Blundell, S. J. & Schofield, A. J. Polarized neutron diffraction of magnetic structures. *Phys. Rev. B* **46**, 3391–3400 (1992).

33. Majkrzak, C. F. & O'Donovan, K. V. Phase-sensitive neutron scattering in thin magnetic films. *Phys. Rev. A* **89**, 33851 (2014).

34. Majkrzak, C. F., Berk, N. F. & Maranville, B. B. Neutron reflectometry for materials science. *J. Appl. Crystallogr.* **55**, 787–812 (2022).

35. Ott, F. & Kozhevnikov, S. Off-specular data representations in neutron reflectivity. *J. Appl. Crystallogr.* **44**, 359–369 (2011).

36. Dorner, B. & Wildes, A. R. Some considerations on resolution and coherence length in reflectometry. *Langmuir* **19**, 7823–7828 (2003).

37. Wildes, A. R. Scientific Reviews: Neutron Polarization Analysis Corrections Made Easy. *https://doi.org/10.1080/10448630600668738* **17**, 17–25 (2007).

38. Pospelov, G. *et al.* BornAgain: software for simulating and fitting grazing-incidence small-angle scattering. *J. Appl. Crystallogr.* **53**, 262–276 (2020).

39. Prokscha, T. *et al.* The new μ E 4 beam at PSI: A hybrid-type large acceptance channel for the generation of a high intensity surface-muon beam. *Nucl. Instruments Methods Phys. Res. Sect. A Accel. Spectrometers, Detect. Assoc. Equip.* **595**, 317–331 (2008).

40. Suter, A. & Wojek, B. M. Musrfit: A Free Platform-Independent Framework for μsR Data Analysis. *Phys. Procedia* **30**, 69–73 (2012).

41. Frąckowiak, Ł. *et al.* Magnetic Domains without Domain Walls: A Unique Effect of He^{+} Ion

Bombardment in Ferrimagnetic Tb/Co Films. *Phys. Rev. Lett.* **124**, 47203 (2019).

42. Lauter, V., Lauter, H. J. C., Glavic, A. & Toperverg, B. P. *Reflectivity, Off-Specular Scattering, and GISANS Neutrons*. *Reference Module in Materials Science and Materials Engineering* (Elsevier Ltd., 2016). doi:10.1016/b978-0-12-803581-8.01324-2.

43. Chien, C. L. & Unruh, K. M. Comparison of amorphous and crystalline FeB. *Phys. Rev. B* **29**, 207 (1984).

44. Clemens, B. M. & Neumeier, J. J. Ion-beam-mixed iron boron films. *J. Appl. Phys.* **58**, 4061–4064 (1985).

45. Platt, C. L., Minor, N. K. & Klemmer, T. J. Magnetic and structural properties of FeCoB thin films. *IEEE Trans. Magn.* **37**, 2302–2304 (2001).

46. Steiner, R. *et al.* Interface reactions in [Fe/B]n multilayers: a way to tune from crystalline/amorphous layer sequences to homogeneous amorphous FexB100-x films. *Appl. Phys. A* **76**, 5–13 (2003).

47. Jackson, T. J. *et al.* Depth-Resolved Profile of the Magnetic Field beneath the Surface of a Superconductor with a Few nm Resolution. *Phys. Rev. Lett.* **84**, 4958–4961 (2000).